\documentclass[10pt,twocolumn,letterpaper]{article}

\usepackage{cvpr}              

\usepackage{graphicx}
\usepackage{amsmath}
\usepackage{amssymb}
\usepackage{booktabs}

\usepackage[pagebackref,breaklinks,colorlinks]{hyperref}
\usepackage{multirow}
\usepackage{amsthm}
\theoremstyle{plain}
\newtheorem{prop}{Proposition}
\theoremstyle{definition}

\DeclareMathOperator*{\argmin}{\arg\min}

\usepackage[capitalize]{cleveref}
\crefname{section}{Sec.}{Secs.}
\Crefname{section}{Section}{Sections}
\Crefname{table}{Table}{Tables}
\crefname{table}{Tab.}{Tabs.}

\def\cvprPaperID{11018} 
\def\confName{CVPR}
\def\confYear{2023}

\begin{document}

\title{DINER: Disorder-Invariant Implicit Neural Representation}

\author{Shaowen Xie$^{1,\dagger}$, Hao Zhu$^{1,\dagger}$, Zhen Liu$^{1,\dagger}$, Qi Zhang$^2$, You Zhou$^1$, Xun Cao$^{1,*}$, Zhan Ma$^{1,*}$\\
$^1$ School of Electronic Science and Engineering, Nanjing University, Nanjing 210023, China\\
$^2$ AI Lab, Tencent Company, Shenzhen 518054, China\\
$\dagger$ Equal contribution.
$^*$ Corresponding authors: {\tt \{caoxun, mazhan\}@nju.edu.cn}
}
\maketitle

\begin{abstract}
Implicit neural representation (INR) characterizes the attributes of a signal as a function of corresponding coordinates which emerges as  a sharp weapon for solving inverse problems. However, the capacity of INR is limited by the spectral bias in the network training. In this paper, we find that such a  frequency-related problem could be largely solved by re-arranging the coordinates of the input signal, for which we propose the disorder-invariant implicit neural representation (DINER) by augmenting a hash-table to a traditional INR backbone. Given discrete signals sharing the same histogram of attributes and different arrangement orders, the hash-table could project the coordinates into the same distribution for which the mapped signal can be better modeled using the subsequent INR network, leading to significantly alleviated spectral bias. Experiments not only reveal the generalization of the DINER for different INR backbones (MLP vs. SIREN)  and  various tasks (image/video representation, phase retrieval, and refractive index recovery) but also show the superiority over the state-of-the-art algorithms both in quality and speed.
\end{abstract}

\section{Introduction}
\label{sec:intro}

INR~\cite{sitzmann2020implicit} continuously describes a signal, providing the advantages of Nyquist-sampling-free scaling, interpolation, and extrapolation without requiring the storage of additional samples~\cite{martel2021acorn}. By combining it with differentiable physical mechanisms such as the ray-marching rendering~\cite{mildenhall2020nerf, kellnhofer2021neural}, Fresnel diffraction propagation~\cite{zhu2022dnf} and partial differential equations~\cite{karniadakis2021physics}, INR becomes a universal and sharp weapon for solving inverse problems and has achieved significant progress in various scientific tasks, \textit{e.g.}, the novel view synthesis~\cite{tewari2022advances}, intensity diffraction tomography~\cite{liu2022recovery} and multiphysics simulation~\cite{karniadakis2021physics}. 


\begin{figure}[t]
  \centering
  \includegraphics[width=0.8\linewidth]{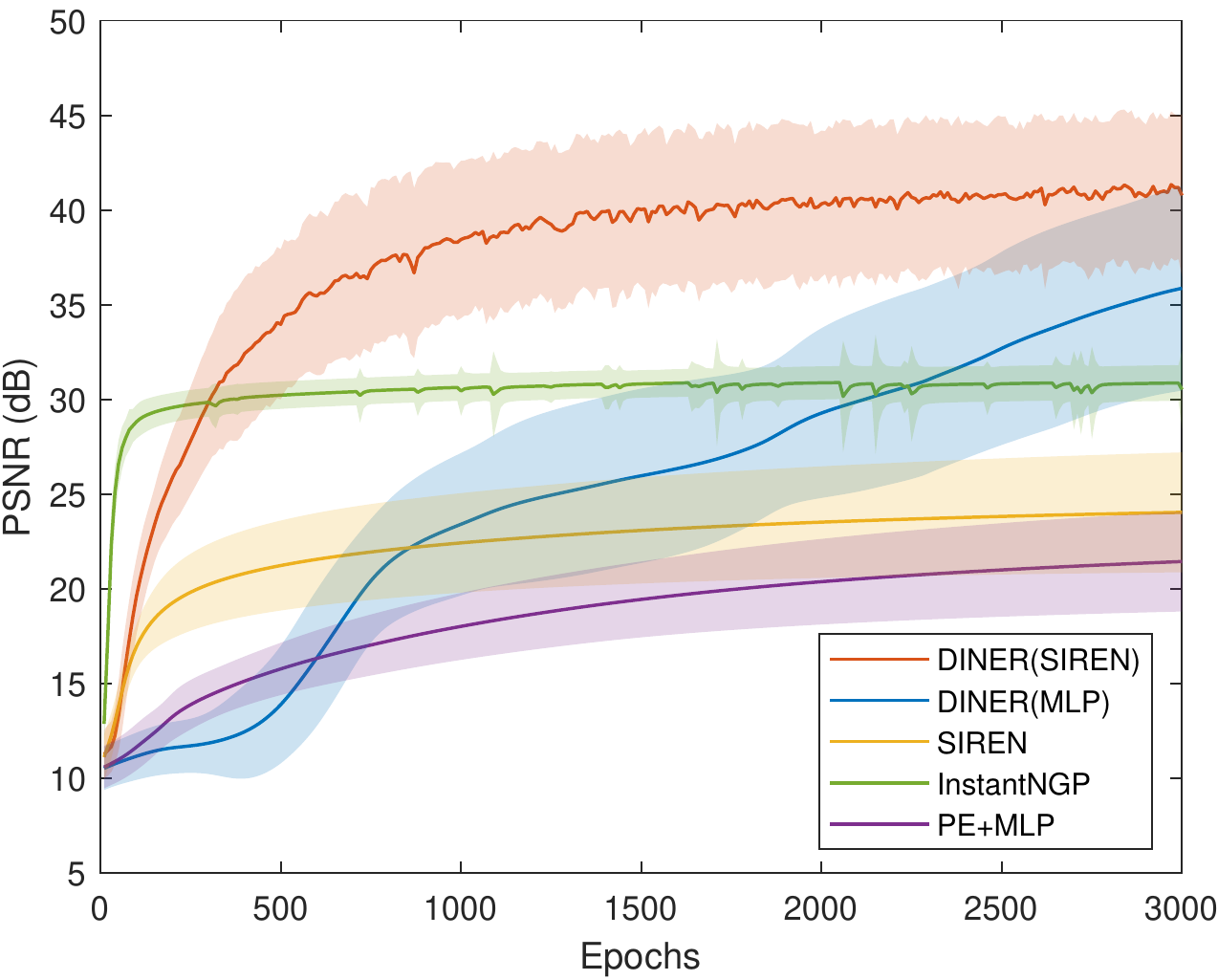}
  \caption{PSNR of various INRs on 2D image fitting over different training epochs.}
  \label{fig:PSNR_over_epochs}
\end{figure}

\begin{figure*}[t]
  \centering
  \includegraphics[width=\linewidth]{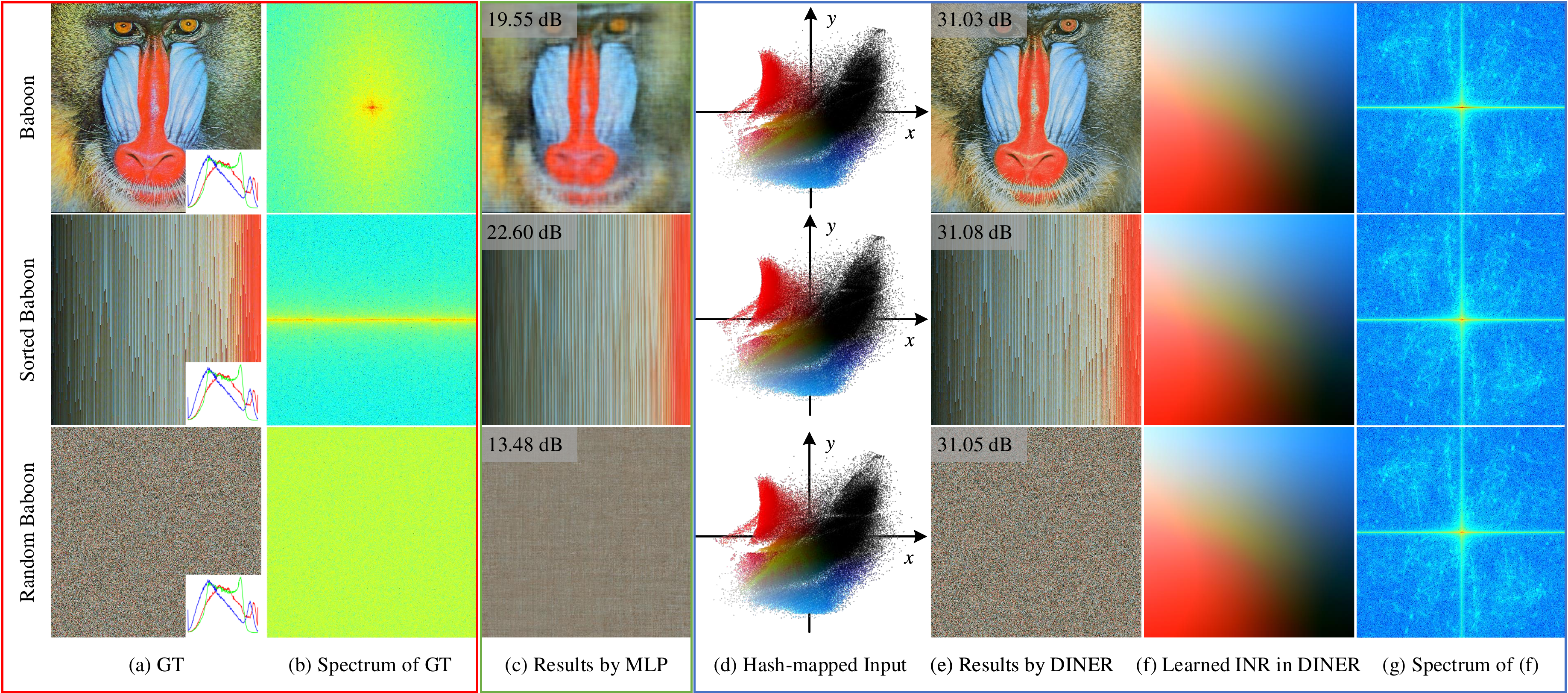}
  \caption{Comparisons of the existing INR and DINER for representing Baboon with different arrangements. From top to bottom, pixels in the Baboon are arranged in different geometric orders, while the histogram is not changed (the right-bottom panel in (a)). From left to right, (a) and (b) refer to the ground truth image and its Fourier spectrum. (c) contains results by an MLP with positional encoding (PE+MLP)~\cite{tancik2020fourier} at a size of $2\times 64$. (d) refers to the hash-mapped coordinates. (e) refers to the results of the DINER that uses the same-size MLP as (c). (f) refers to the learned INR in DINER by directly feeding grid coordinates to the trained MLP (see Sec.~\ref{sec:method_detail} for more details). (g) is the Fourier spectrum of (f). (g) shares the same scale bar with (b).}
  \label{fig:freq_cmp_rgb}
\end{figure*}

However, the capacity of INR is often limited by the underlying network model itself. For example, the spectral bias~\cite{rahaman2019spectral} usually makes the INR easier to represent low-frequency signal components (see Fig.~\ref{fig:PSNR_over_epochs} and Fig.~\ref{fig:freq_cmp_rgb}(c)). To improve the representation capacity of the INR model, previous explorations mainly focus on encoding more frequency bases using either Fourier basis~\cite{mildenhall2020nerf, tancik2020fourier, sitzmann2020implicit, yuce2022structured} or wavelet basis~\cite{fathony2020multiplicative, lindell2021bacon} into the network. However, the length of function expansion is infinite in theory, and a larger model with more frequency bases runs exceptionally slowly.

Such a problem is closely related to the input signal's frequency spectrum. The signal's frequency tells how fast the signal attribute changes following the intrinsic order of geometry coordinates. By properly re-arranging the order of coordinates of a discrete signal, we could modulate the signal's frequency spectrum to possess more low-frequency components. Such a re-arranged signal can be better modeled using subsequent INR. Based on this observation, we propose the DINER. In DINER, the input coordinate is first mapped to another index using a hash-table and fed into a traditional INR backbone. We prove that no matter what orders the elements in the signal are arranged, the joint optimization of the hash-table and the network parameters guarantees the same mapped signal (Fig.~\ref{fig:freq_cmp_rgb}(d)) with more low-frequency components. As a result, the representation capacity of the existing INR backbones and the task performance are largely improved. As in Fig.~\ref{fig:PSNR_over_epochs} and Fig.~\ref{fig:freq_cmp_rgb}(e), a tiny shallow and narrow MLP-based INR could well characterize the input signal with arbitrary arrangement orders. The use of hash-table trades the storage for fast computation where the caching of its learnable parameters for mapping is usually at a similar size as the input signal,  but the computational cost is marginal since the back-propagation for the hash-table derivation is $\mathcal{O}(1)$.

 Main contributions are summarized as follows:
\begin{enumerate}
    \item The inferior representation capacity of the existing INR model is largely increased by the proposed DINER, in which a hash-table is augmented to map the coordinates of the original input for better characterization in the succeeding INR model.
    \item The proposed DINER provides consistent mapping and representation capacity for signals sharing the same histogram of attributes and different arrangement orders.
    \item The proposed DINER is generalized in various tasks, including the representation of 2D images and 3D videos, phase retrieval in lensless imaging, as well as 3D refractive index recovery in intensity diffraction tomography, reporting significant performance gains to the existing state-of-the-arts.
\end{enumerate}


\section{Related Work}
\subsection{INR and inverse problem optimization}

INR (sometimes called coordinate neural network) builds the mapping between the coordinate and its signal value using a neural network, promising continuous and memory-efficient modeling for various signals such as the 1D audio~\cite{gao2022objectfolder}, 2D image~\cite{tancik2020fourier}, 3D shape~\cite{park2019deepsdf}, 4D light field~\cite{sitzmann2021light} and 5D radiance field~\cite{mildenhall2020nerf}. Accurate INR for these signals could be supervised directly by comparing the network output with the ground truth, or an indirect way that calculates the loss between the output after differentiable operators and the variant of the ground truth signal. Thus, INR becomes a universal tool for solving inverse problems because the forward processes in these problems are often well-known. INR has been widely applied in the optimization of inverse problems in several disciplines, such as computer vision and graphics~\cite{tewari2022advances}, computational physics~\cite{karniadakis2021physics}, clinical medicine~\cite{yeung2021implicitvol}, biomedical engineering~\cite{zhu2022dnf, liu2022recovery}, material science~\cite{chen2020physics} and fluid mechanics~\cite{raissi2020hidden, reyes2021learning}.

\subsection{Encoding high-frequency components in INR}
According to the approximation theory, an MLP network could approximate any function~\cite{leshno1993multilayer}. However, there is a spectral bias~\cite{rahaman2019spectral} in the network training, resulting in low performance of INR for high-frequency components. Several attempts have been explored to overcome this bias and could be classified into two categories, \textit{i.e.}, the function expansion and the parametric encodings.

The function expansion idea treats the INR fitting as a function approximation using different bases. Mildenhall et al.~\cite{mildenhall2020nerf} encoded the input coordinates using a series of $\sin / \cos$ functions with different frequencies and achieved great success in radiance field representation. The strategy of frequency-predefined $\sin / \cos$ functions is further improved with random Fourier features, which has been proved to be effective in learning high-frequency components both in theory and in practical~\cite{tancik2020fourier}. Sitzmann et al.~\cite{sitzmann2020implicit} replaced the classical ReLU activation with periodic activation function (SIREN). The different layers of the SIREN could be viewed as increasing different frequency supports of a signal~\cite{yuce2022structured}. SIREN is suited for representing complex natural signals and their derivatives. Apart from the Fourier expansion, Fathony et al.~\cite{fathony2020multiplicative} represented a complex signal by a linear combination of multiple wavelet functions (MFN), where the high-frequency components could be well modeled by modulating the frequency in the Gabor filter. Lindell et al.~\cite{lindell2021bacon} developed MFN and proposed the band-limited coordinate networks, where the frequency at each network layer could be specified at initialization. These methods have achieved significant advantages in representing high-frequency components compared with the standard MLP network. However, \textit{the performance of these INRs are limited by the frequency distribution of a signal-self, and often require a deeper or wider network architecture to improve the fitting accuracy}.

From the perspective of the parameter encoding~\cite{liu2020neural,takikawa2021neural,chabra2020deep,jiang2020local}, each input coordinate is encoded using learned features which are fed into an MLP for fitting. Takikawa et al.~\cite{takikawa2021neural} divided the 3D space using a sparse voxel octree structure where each point is represented using a learnable feature vector from its eight corners, achieving real-time rendering of high-quality signed distance functions. Martel et al.~\cite{martel2021acorn} divided the coordinate space iteratively during the INR training (ACORN), where the encoding features for each local block are obtained by a coordinate encoder network and are fed into a decoder network to obtain the attribute of a signal. ACORN achieves nearly 40 dB PSNR for fitting gigapixel images for the first time. Muller et al.~\cite{muller2022instant} replaced the coordinate encoder network with a multi-resolution hash-table. Because the multi-resolution hash-table has higher freedom for characterizing coordinates' features, only a tiny network is used to map the features and the attribute values of a signal. Despite the superiority of faster convergence and higher accuracy in parameter encoding, one of the key questions is still not answered, \textit{i.e.}, \textit{what are the geometrical meanings of these features?}

Compared with these methods, the hash-table in the proposed DINER unambiguously projects the input coordinate into another, or in other words, mapping the signal into the one with more low-frequency components. As a result, a tiny network could achieve very high accuracy compared with previous methods.


\section{Performance of INR}
\subsection{Background of expressive power of INR}
Following the form of Yuce et al.~\cite{yuce2022structured}, an INR with a 1D input $x$ could be modeled as a function $f_{\theta}(x)$ that maps the input coordinate $x$ to its attribute, that
\begin{equation}
\footnotesize
\begin{aligned}
    \mathbf{z}^{0}&=\gamma (x), \\
    \mathbf{z}^{l}&=\rho^{l}(\mathbf{W}^{l}\mathbf{z}^{l-1}+\mathbf{b}^{l}),\: l=1,...,L-1\\
    f_{\theta}(x)&=\mathbf{W}^{L}\mathbf{z}^{L-1}+\mathbf{b}^{L}
\end{aligned},
\end{equation}
where $\gamma(\cdot)$ is the preprocess function which is often used to encode more frequency bases in the network, $\mathbf{z}^{l}$ is the output of the $l$-th layer in INR, $\rho$ is the activation function, $\mathbf{W}^{l}$ and $\mathbf{b}^l$ are the weight and bias matrix in the $l$-th layer, $L$ is the number of layers in INR, $\theta=\{\mathbf{W}^{l},\mathbf{b}^{l}\}_{1}^{L}$ refers to the set of all training parameters in the network. 

This INR could represent signals with functions following the form
\begin{equation}
\footnotesize
\label{eqn:INR_expressivepower}
f(x)=\sum_{\omega'\in\mathcal{H}_{\Omega}}c_{\omega'}\sin (\langle\omega',x\rangle+\phi_{\omega'}),
\end{equation}
where $\mathcal{H}_{\Omega}$ is the frequency set~\cite{yuce2022structured} determined by the frequency selected in the preprocess function $\gamma(\cdot)$, \textit{e.g.}, the Fourier encoding~\cite{tancik2020fourier}, or the $\sin$ activation~\cite{sitzmann2020implicit}. In other words, \textit{the expressive power of INR is restricted to functions that can be represented using a linear combination of certain harmonics of the $\gamma(\cdot)$}~\cite{yuce2022structured}.

\subsection{Arrangement order of a signal determines the capacity of INR}

According to the expressive power of an INR (Eqn.~\ref{eqn:INR_expressivepower}), a signal could be well learned when the encoded frequencies in the INR are consistent with the signal's frequency distribution. However, there are two problems in applying this conclusion,
\begin{enumerate}
    \item The frequency distribution of a signal could not be known in advance, especially in inverse problems. Thus proper frequencies could not be well set in designing the architecture of an INR.
    \item Due to the spectral bias in network training~\cite{rahaman2019spectral}, the low-frequency components in a signal will be learned first, while the high-frequency components are learned in an extremely slow convergence~\cite{ronen2019convergence, bietti2019inductive, heckel2020compressive, tancik2020fourier}.
\end{enumerate}

We notice that most of the signals recorded or to be inversely solved today are discrete signals. The frequency distribution of a discrete signal could be changed by arranging elements in different orders at the cost of additional storage for the arrangement rule, resulting in different satisfactions of Eqn.~\ref{eqn:INR_expressivepower}. Consequently, the capacity of an INR for representing a signal changes with different arrangement orders.

Fig.~\ref{fig:freq_cmp_rgb} gives an intuitive demonstration. The Baboon\footnote{Baboon is a self-contained image in the Matlab by MathWorks$^\copyright$.} image is arranged in different orders in Fig.~\ref{fig:freq_cmp_rgb}(a), and the corresponding Fourier spectrum images are shown in Fig.~\ref{fig:freq_cmp_rgb}(b). The original image contains rich low-, intermediate- and high-frequency information. By sorting the Baboon according to the intensities of pixels, the high-frequency information in $y$-axis almost disappeared. We then arrange the pixels using a random order. Currently, the Baboon contains much high-frequency information. Then a PE+MLP ($2\times 64$, \textit{i.e.}, 2 hidden layers and 64 neurons per layer with ReLU activation) network is applied to learn the mapping between the coordinates and intensities of these three images (Fig.~\ref{fig:freq_cmp_rgb}(c)). The fitting results differ significantly. The PE+MLP gets the best performance in the sorted image, which contains the most low-frequency information, while the worst results appear in the random sorted image, which contains the most high-frequency information. In summary:

\begin{prop}
\label{prop_signal_order}
Different arrangements of a signal have different frequency distributions, resulting in different capacities of INR for representing the signal-self.
\end{prop}

\section{Disorder-invariant INR}

\subsection{Hash-mapping for INR}
\label{sec:method_detail}
Given a paired discrete signal $Y=\{(\vec{x}_i,\vec{y}_i)\}_{i=1}^{N}$, where $\vec{x}_i$ be the $i$-th $d_{in}$-dimensional coordinate, and $\vec{y}_i$ be the corresponding $d_{out}$-dimensional signal attribute. Following the analysis mentioned above, an ideal arrangement rule $M^*:\:\mathbb{R}^{d_{in}}\rightarrow \mathbb{R}^{d_{in}}$ should meets the following rule,
\begin{subequations}
\footnotesize
\label{eqn:condition_best_mapping}
\begin{align}
      {M}^{*} &= \argmin_{M}\sum_{k=1}^{K_M}|\omega_{k}|      \label{eqn:condition_best_mapping:a}\\
      \Omega_{M} &\subseteq  \mathcal{H}_{\Omega},\:\:
      \Omega_{M} =\{\omega_{k}\}_{k=1}^{K_{M}},  \label{eqn:condition_best_mapping:b}
\end{align}
\end{subequations}
where $\Omega_{M}$ is the set of frequency by mapping the signal following the rule $M$, $\mathcal{H}_{\Omega}$ is the supported frequency set of the INR network (Eqn.~\ref{eqn:INR_expressivepower}), $K_{M}$ is the number of frequency in the arranged signal, $|\cdot|$ returns the absolute value of $\cdot$. The signal with this arrangement could be well learned since both the problems of improper frequency setting (Eqn.~\ref{eqn:condition_best_mapping:b}) and the spectral bias (Eqn.~\ref{eqn:condition_best_mapping:a}) are taken into account.

However, this strategy requires prior knowledge of the signal distribution, which is only suitable for the compression task. In contrast, it losses the ability to optimize inverse problems where the signal distribution to be optimized could not be achieved in advance. In this subsection, we detail the proposed DINER to handle this problem.

\begin{figure}[t]
  \centering
  \includegraphics[width=\linewidth]{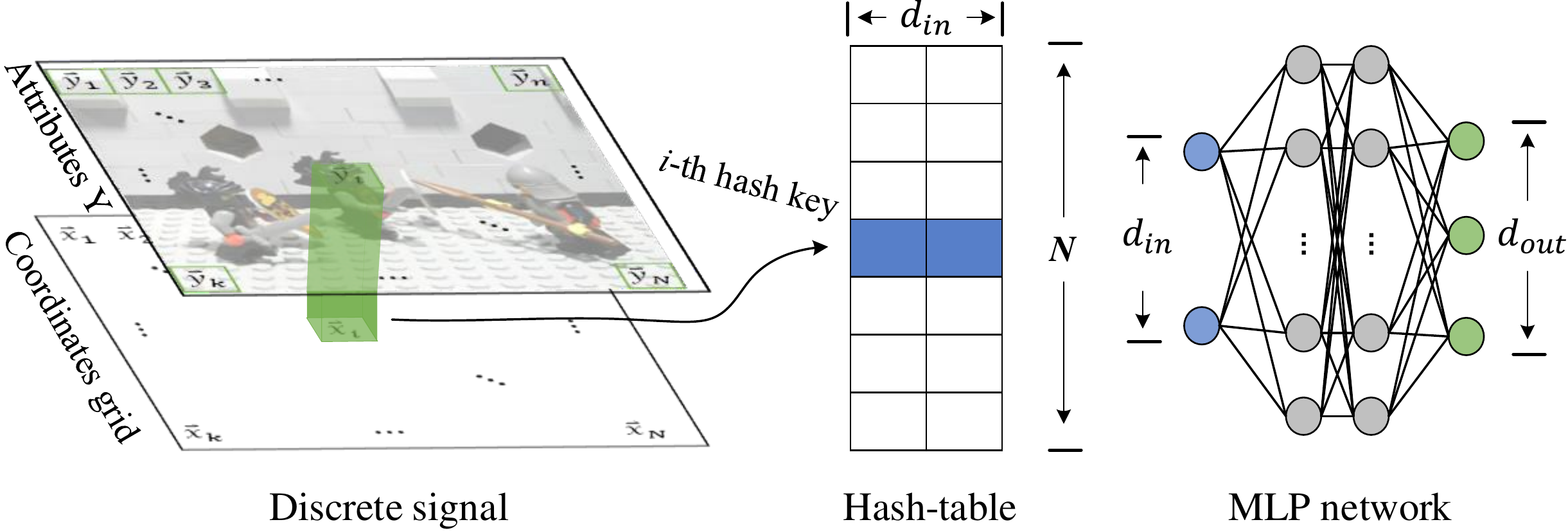}
  \caption{Pipeline of the DINER.}
  \label{fig:hash_table_INR_demo}
\end{figure}

We specifically design a full-resolution hash-table $\mathcal{HM}$ to model the mapping mentioned above. The hash-table $\mathcal{HM}$ is set as the same length $N$ as the number of elements in $Y$, meanwhile the width of $\mathcal{HM}$ is set as $d_{in}$ (the dimensions of $\vec{x}_i$). Firstly, the input coordinate $\vec{x}_i$ is used to query the $i$-th hash key $M(\vec{x}_i)$ in the $\mathcal{HM}$. Then the mapped coordinate $M(\vec{x}_i)$ is fed into a standard MLP. All parameters in the $\mathcal{HM}$ are set as learnable, \textit{i.e.}, parameters in $\mathcal{HM}$ and the network parameters will be jointly optimized during the training process. Fig.~\ref{fig:hash_table_INR_demo} demonstrates the above process (The Lego Knight used comes from \cite{stanford_lf_web}.).

Due to the hash-table, the MLP network actually learns the mapped signal. Fig.~\ref{fig:freq_cmp_rgb}(d) shows the mapped pixels of the Baboon after hash-table, it is noticed that the original grid coordinates (Fig.~\ref{fig:freq_cmp_rgb}(a)) are projected into irregular points (Fig.~\ref{fig:freq_cmp_rgb}(d)). We sample a mesh evenly according to the minimum and the maximum values in the mapped coordinates, and feed them into the trained MLP, \textit{i.e.}, the Fig.~\ref{fig:freq_cmp_rgb}(f). For simplicity, the image in Fig.~\ref{fig:freq_cmp_rgb}(f) is later called `learned INR'. The learned INR differs significantly from the Baboon in that the former is much smoother than the latter and has many low-frequency components (Fig.~\ref{fig:freq_cmp_rgb}(g)). As a result, high accuracy of MLP fitting could be achieved using the hash-table (Fig.~\ref{fig:freq_cmp_rgb}(e)).

\subsection{Analysis of disorder-invariance}
When applying the INR to tasks of signal representation and inverse problem optimization, the training or the optimization of the traditional INR and DINER could be modeled as Eqns.~\ref{eqn:loss_func_cmp:CON} and~\ref{eqn:loss_func_cmp:Diner}, respectively.

\begin{subequations}
\footnotesize
\label{eqn:loss_func_cmp}
\begin{align}
    \theta^*
    &=\argmin_{\theta}\mathcal{L}\left(\mathcal{P}\left(\{f_{\theta}(\vec{x}_i)\}_{1}^{N}\right), \mathcal{P}\left(\{\vec{y}_i\}_{1}^{N}\right)\right) \label{eqn:loss_func_cmp:CON}\\
    \theta^*,\mathcal{HM}^*
    &=\argmin_{\theta,\mathcal{HM}}\mathcal{L}\left(\mathcal{P}\left(\{f_{\theta}(\mathcal{HM}(\vec{x}_i))\}_{1}^{N}\right), \mathcal{P}\left(\{\vec{y}_i\}_{1}^{N}\right)\right)\nonumber\\
    &=\argmin_{\theta,\mathcal{HM}}\mathcal{L}\left(\mathcal{P}\left(\{f_{\theta}(\mathcal{HM}_i)\}_{1}^{N}\right), \mathcal{P}\left(\{\vec{y}_i\}_{1}^{N}\right)\right),\label{eqn:loss_func_cmp:Diner}
\end{align}
\end{subequations}
where $\mathcal{P}$ is a physical process and is an identical transformation for the representation task, $\mathcal{L}$ is the loss function according to the measurements and the reconstructed results, $\mathcal{HM}_i$ is the $i$-th key in the hash-table. Because the hash-index operation has no gradient, the first equation in Eqn.~\ref{eqn:loss_func_cmp:Diner} could be simplified to the second one in Eqn.~\ref{eqn:loss_func_cmp:Diner}.

It is noticed that paired relationship between the coordinate $\vec{x}_i$ and value $\vec{y}_i$ in Eqn.~\ref{eqn:loss_func_cmp:CON} is broken in the Eqn.~\ref{eqn:loss_func_cmp:Diner}. There is only one independent variable $\vec{y}_i$ in the loss function Eqn.~\ref{eqn:loss_func_cmp:Diner}. Assuming parameters in $\theta$ are initialized with the same values and all keys in $\mathcal{HM}$ are set with the same one ($\textit{e.g.}$, 0) in every experiment, when applying a different order to the signal $Y$, \textit{e.g.}, $Y'=\{\vec{x}_j,\vec{y}_i\}_{j=N,i=1}^{j=1,i=N}$, the training of $\theta$ and $\mathcal{HM}$ for signals $Y$ and $Y'$ share the same optimization progress in every gradient update since all parameters in Eqn.~\ref{eqn:loss_func_cmp:Diner} for $Y$ are equivalent to ones for $Y'$. As a result, the same $\theta^*$ will be optimized while the $\mathcal{HM}'$ of $Y'$ is also an inverse arrangement of the $\mathcal{HM}$ of $Y$. This equivalence is not limited to the $Y'$ with an inverse order; actually, it could be easily proved that the equivalence holds for $Y$ with an arbitrary order. 

Fig.~\ref{fig:freq_cmp_rgb}(d), (e), and (f) illustrate this equivalence. Although the Baboon is arranged with different orders, the hash-table maps them into the same signal (Fig. \ref{fig:freq_cmp_rgb}(d), (f)), and DINER optimizes them with similar PSNR values $31.03,\: 31.08,\: 31.05$\footnote{The slight difference in values comes from the floating point errors of GPU for summing matrix with same histogram and different arrangement orders.}. In summary:

\begin{prop}
\label{prop_disorder_invariance}
The DINER is disorder-invariant, and signals with the same histogram distribution of attributes share an optimized network with the same parameter values.
\end{prop}

\subsection{Discussion}
\noindent \textbf{Backbone network.} The backbone of the proposed DINER is not limited to the standard MLP used above; actually, other network structures such as the SIREN could also be integrated with the hash-table and get better performance than the original structure. Please refer to the experimental section for more details.

\noindent \textbf{Complexity.}  Although the number of parameters of the hash-table is much larger than the network, the training cost is very small because only one hash-key needs to be updated for training an MLP with batchsize 1. As a result, the computational complexity of training hash-table is $\mathcal{O}(1)$ in each iteration of training.



\section{Experiments}
To verify the performance of the proposed DINER, four separate tasks are conducted: 2D image fitting, neural representation for 3D video, phase retrieval in lensless imaging, and 3D Refractive Index recovery in intensity diffraction tomography.

\subsection{2D Image Fitting}
\subsubsection{Dataset and Algorithm Setup}
The 2D image fitting task is adopted to test the performance of the proposed DINER and to illustrate the change of frequency distribution of the signal. We use 30 high-resolution images with $1200\times 1200$ resolution from the SAMPLING category of the TESTIMAGES dataset~\cite{asuni2014testimages} to evaluate the performance of various algorithms. Each image is generated using custom Octave/MATLAB software scripts specifically written to guarantee the precise positioning and value of every pixel and contains rich low-, intermediate- and high-frequency information. In the following experiments, all our results are obtained with the same network configuration $2\times 64$ unless otherwise stated.

\begin{table}
\footnotesize
  \centering
  \begin{tabular}{@{}ccccc@{}}
    \toprule
    Freq. bands & \begin{minipage}{0.07\textwidth}
      \includegraphics[width=10mm, height=10mm]{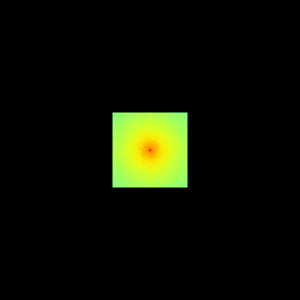}
    \end{minipage} & 
    \begin{minipage}{0.07\textwidth}
      \includegraphics[width=10mm, height=10mm]{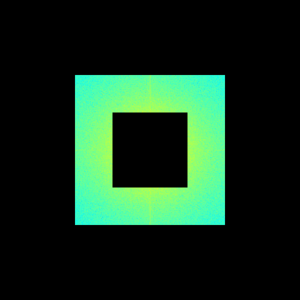}
    \end{minipage} & 
    \begin{minipage}{0.07\textwidth}
      \includegraphics[width=10mm, height=10mm]{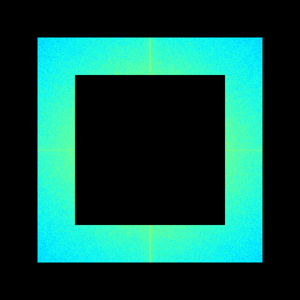}
    \end{minipage} &
    \begin{minipage}{0.07\textwidth}
      \includegraphics[width=10mm, height=10mm]{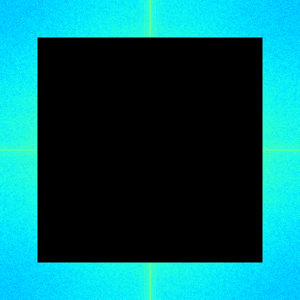}
    \end{minipage}\\
    \midrule
    Original Image & 0.4426 & 0.2484 & 0.1753 & 0.1337 \\
    Learned INR  & \multirow{2}{*}{0.6784} & \multirow{2}{*}{0.1218} & \multirow{2}{*}{0.0973} & \multirow{2}{*}{0.1025} \\
    (MLP) & & & & \\
    Learned INR & \multirow{2}{*}{0.6354} & \multirow{2}{*}{0.1220} & \multirow{2}{*}{0.1149} & \multirow{2}{*}{0.1276} \\
    (SIREN) & & & &\\
    \bottomrule
  \end{tabular}
  \caption{Comparisons of the ratio of frequency distributions between the original image and the learned INR. Two backbones, \textit{i.e.}, the MLP and SIREN, are both compared.}
  \label{tab:freq_dist_w_wo_hash}
\end{table}

\begin{figure}[t]
  \centering
  \includegraphics[width=\linewidth]{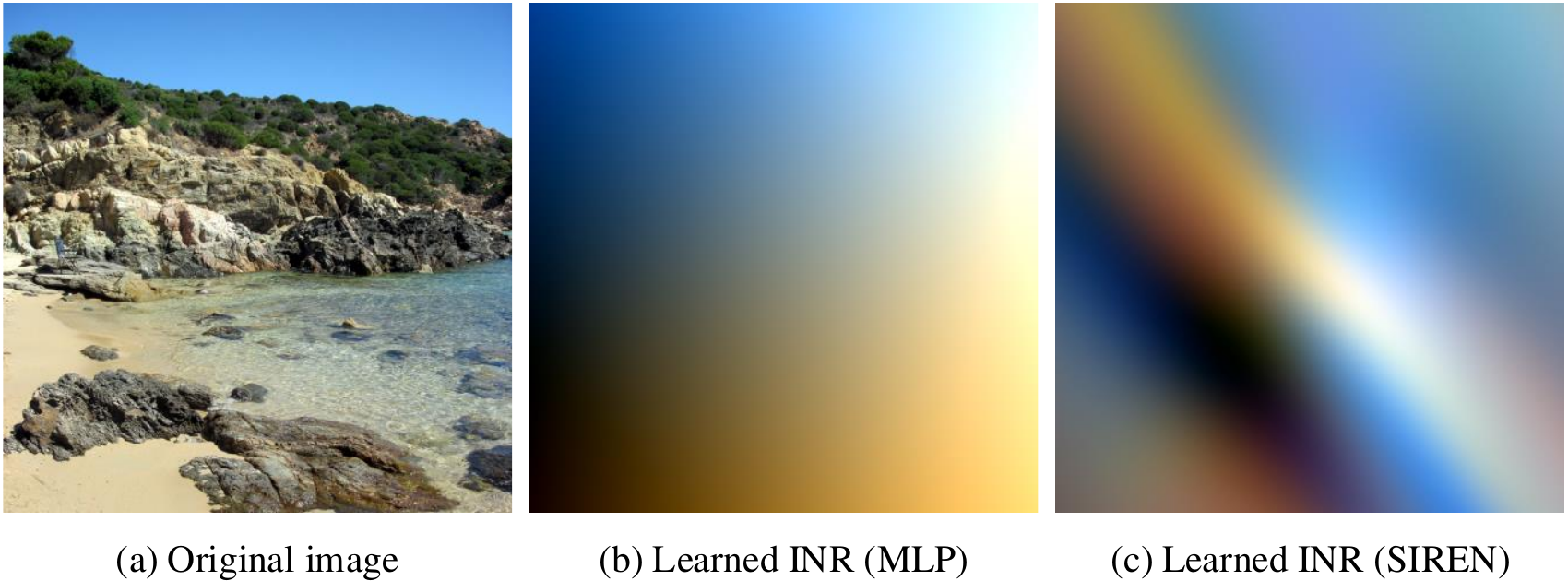}
  \caption{Comparisons of learned INRs in the DINER with MLP and SIREN backbones, respectively.}
  \label{fig:image_cmp_siren_mlp}
\end{figure}

\begin{figure*}[t]
  \centering
  \includegraphics[width=0.9\linewidth]{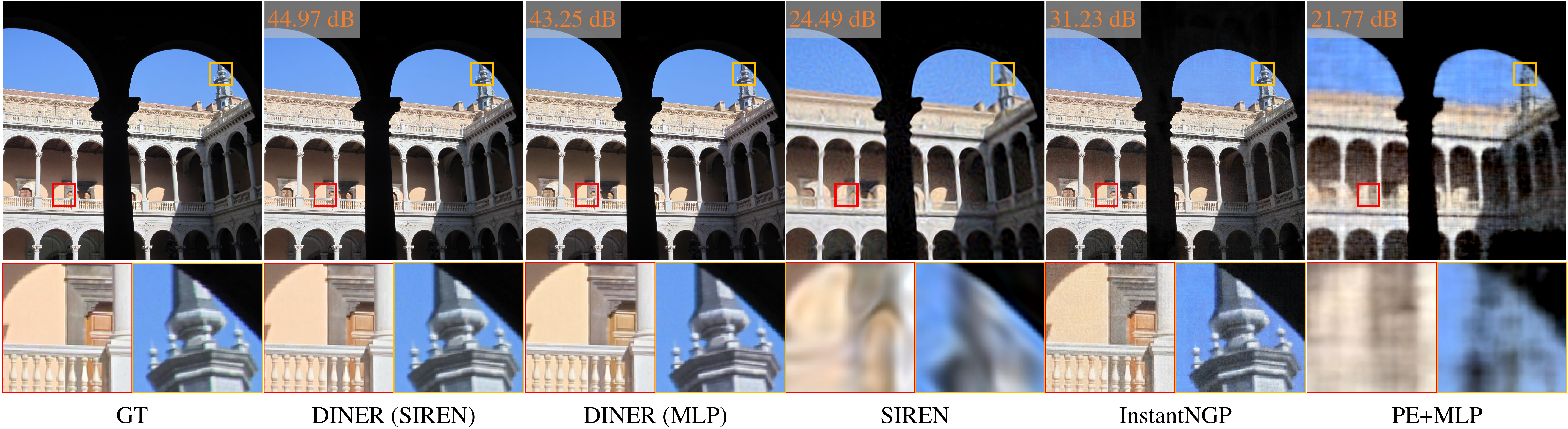}
  \caption{Qualitative comparisons of various methods on 2D image fitting after 3000 epochs.}
  \label{fig:image_res_cmp}
\end{figure*}

\subsubsection{Comparisons of frequency distribution with and without hash-table}
As noticed in the Sec.~\ref{sec:method_detail} and the Figs.~\ref{fig:freq_cmp_rgb}, the hash-table maps the original signal with more low-frequency contents. Tab.~\ref{tab:freq_dist_w_wo_hash} provides statistics of the mean frequency distributions of the original image and the learned INR over 30 images. The ratios of the intermediate- and high-frequency information are all reduced after mapping, while the low-frequency information are increased. 

Compared with the supported frequency set $\mathcal{H}_{\Omega}^{SIREN}$ of the SIREN structure where a default frequency $30$ is used in activation, the $\mathcal{H}_{\Omega}^{MLP}$ of the MLP structure without any frequency encoding contains more low frequencies and less high frequencies. Accordingly, there are more low-frequency information in the mapped image of DINER with MLP backbone than the one with SIREN backbone ($0.6784$ \textit{vs} $0.6354$, Fig.~\ref{fig:image_cmp_siren_mlp}). Please refer to the supplementary material for more qualitative comparisons. 

\subsubsection{Comparisons with the State-of-the-arts}
We compare the proposed DINER with the Fourier feature positional encoding (PE+MLP)~\cite{tancik2020fourier}, SIREN~\cite{sitzmann2020implicit} and InstantNGP~\cite{muller2022instant}. Noting that, two backbones, \textit{i.e.}, the standard MLP with ReLU activation and the SIREN with periodic function activation, are all combined with the proposed hash-table to better evaluate the performance. We control the size of the hash-table used in the InstantNGP to guarantee the similar parameters with ours, \textit{e.g.}, $2^{21}$ in the 2D image fitting task while ours has a length of $1200^2 < 2^{21}$. Apart from this, all 5 methods are trained with the same $L_{2}$ loss between the predicted value and the ground truth, and other parameters are set with the default values by authors.

Fig.~\ref{fig:PSNR_over_epochs} shows the PSNR of various methods at different epochs. It is noticed that the SIREN and PE+MLP convergences quickly at the early stage and reaches about $24$dB and $21$dB finally. On the contrary, the proposed two methods both provide higher accuracy than backbones. The PSNRs of two backbones for image fitting are increased $14$dB and $16$dB using the hash-table, respectively. Additionally, although the InstantNGP converges very fast to about $30$dB at about 200 epochs, the curve tends to be stable at the last 2800 epochs. The proposed DINER with MLP backbone achieves an advantage of $5$dB than the InstantNGP.

Fig.~\ref{fig:image_res_cmp} shows the qualitative results at 3000 epochs. The proposed methods outperform the SIREN and PE+MLP. Our methods provide more clear details especially in high-frequency boundaries, such as the bell tower (yellow box) in Fig.~\ref{fig:image_res_cmp}. The fitted image of the InstantNGP is very similar to the GT at first sight, however many noises appear in the zoom-in results, for example there are a lot of noisy points in the wall (red box) and the sky (yellow box) of Fig.~\ref{fig:image_res_cmp}, results in a lower PSNR metric.

Tab.~\ref{tab:time_image} lists the training time of 5 methods. The InstantNGP is implemented with the tiny-cuda-nn~\cite{muller2022tinycudann}, while other 4 methods are implemented with the Pytorch. All 5 methods are trained on a NVIDIA A100 40GB GPU. The optimization of hash-table requires additional 3 seconds on the SIREN architecture and reduces 20 seconds compared with the classical PE+MLP architecture, verifying the low complexity of optimizing hash-table.

\begin{table}
\footnotesize
  \centering
  \begin{tabular}{@{}lccccc@{}}
    \toprule
     & DINER & DINER & \multirow{2}{*}{SIREN} & \multirow{2}{*}{InstantNGP} & PE \\
     & (SIREN) & (MLP) & & &+MLP\\
    \midrule
    Time& 80.5s & 58.5s & 77.6s & 38s &78.8s \\
    \bottomrule
  \end{tabular}
  \caption{Comparisons of training a 2D image with 3000 epochs.}
  \label{tab:time_image}
\end{table}

\begin{figure}[t]
  \centering
  \includegraphics[width=\linewidth]{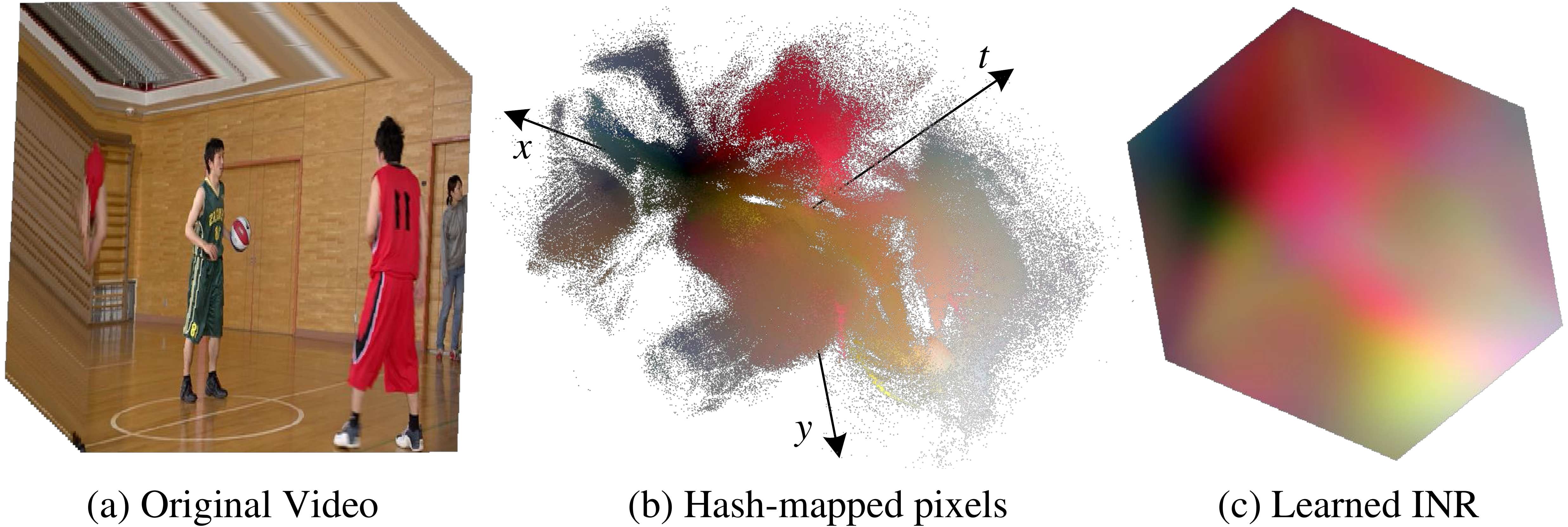}
  \caption{Visualization of learned INR on the 3D video `BasketballPass'~\cite{sullivan2012overview}. (a) and (b) compare the coordinates with and without hash-table. (c) shows the learned INR of the SIREN after the mapping by hash-table.}
  \label{fig:coords_mapped_video}
\end{figure}

\begin{figure*}[t]
  \centering
  \includegraphics[width=0.9\linewidth]{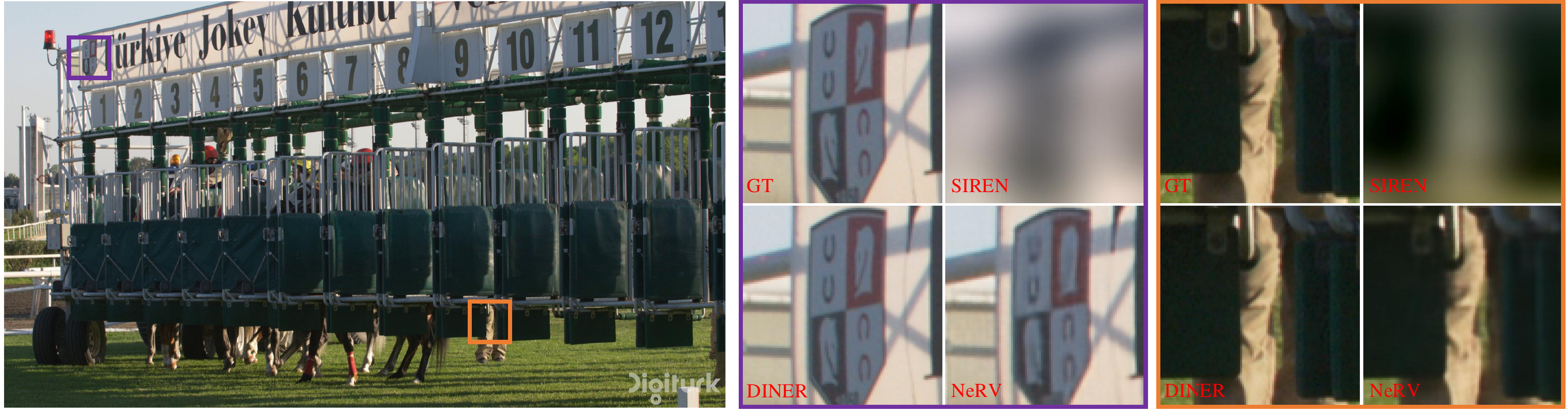}
  \caption{Qualitative comparisons of various methods on 3D video representation after 500 epochs.}
  \label{fig:video_res_cmp}
\end{figure*}

\subsection{Neural Representation for 3D video}
Video describes a dynamic 3D scene $I(t,x,y)$ composed of multiple frames. Accurate representation for 3D video is becoming a popular task~\cite{chen2021nerv,rho2022neural,kim2022scalable,li2022nerv} in the community of INR. We compare the proposed DINER with the SIREN~\cite{sitzmann2020implicit} and the state-of-the-art INR NeRV~\cite{chen2021nerv}. Noting only the Hash+SIREN architecture is evaluated in this task. The NeRV is implemented using their default parameters, while the SIREN and the proposed Hash+SIREN both use the same network structure with the size $4\times 64$. All three methods are evaluated on the videos 'ReadySetGo' and 'ShakeNDry' of the UVG dataset~\cite{mercat2020uvg} and are trained with 500 epochs. The first 30 frames with 1920$\times$1080 resolution are used in our experiment.

\begin{table}
\small
  \centering
  \begin{tabular}{@{}lccc@{}}
    \toprule
    Methods & Network Para. & Training time & PSNR  \\
    \midrule
    SIREN & 8.77K & 706s & 21.22 dB \\
    NeRV & 97.24M & 7445s & 36.08 dB\\
    DINER & 8.77K & 1309s & 50.74 dB \\
    \bottomrule
  \end{tabular}
  \caption{Comparisons of training 3D videos 'ReadySetGo' and 'ShakeNDry' with 500 epochs.}
  \label{tab:3D_video_psnr}
\end{table}

Fig.~\ref{fig:coords_mapped_video}(a) and (b) show the mapping of the coordinates with and without mapping. Fig.~\ref{fig:coords_mapped_video}(c) illustrates the learned INR. It is noticed that the low-frequency property also appears in the learned INR of the 3D video in our DINER. Tab.~\ref{tab:3D_video_psnr} shows the quantitative comparisons. The proposed method outperforms the NeRV both in quality and speed with $14$ dB and $5\times$ improvements, respectively. Because a large hash-table is used in DINER, more time are taken in the transmission between the memory and the cache in the GPU, resulting more training time of DINER than the SIREN. Fig.~\ref{fig:video_res_cmp} shows the qualitative comparisons. Noting that the SIREN with a tiny network could not provide reasonable representation for the `ReadySetGo' data with $30\times 1920\times 1080$ pixels, thus all pixels are smoothed. NeRV provides better results than the ones by SIREN, however the high-frequency details are lost such as the character `U' (left-top corner) and the red logo of horsehead (right-top corner) in the purple box, as well as the folds of the trousers in the orange box. On the contrary, the original video is mapped with little high-frequency component in the proposed DINER (Fig.~\ref{fig:coords_mapped_video}(c)). As a result, the details mentioned above could be well represented.

\begin{figure}[t]
  \centering
  \includegraphics[width=0.8\linewidth]{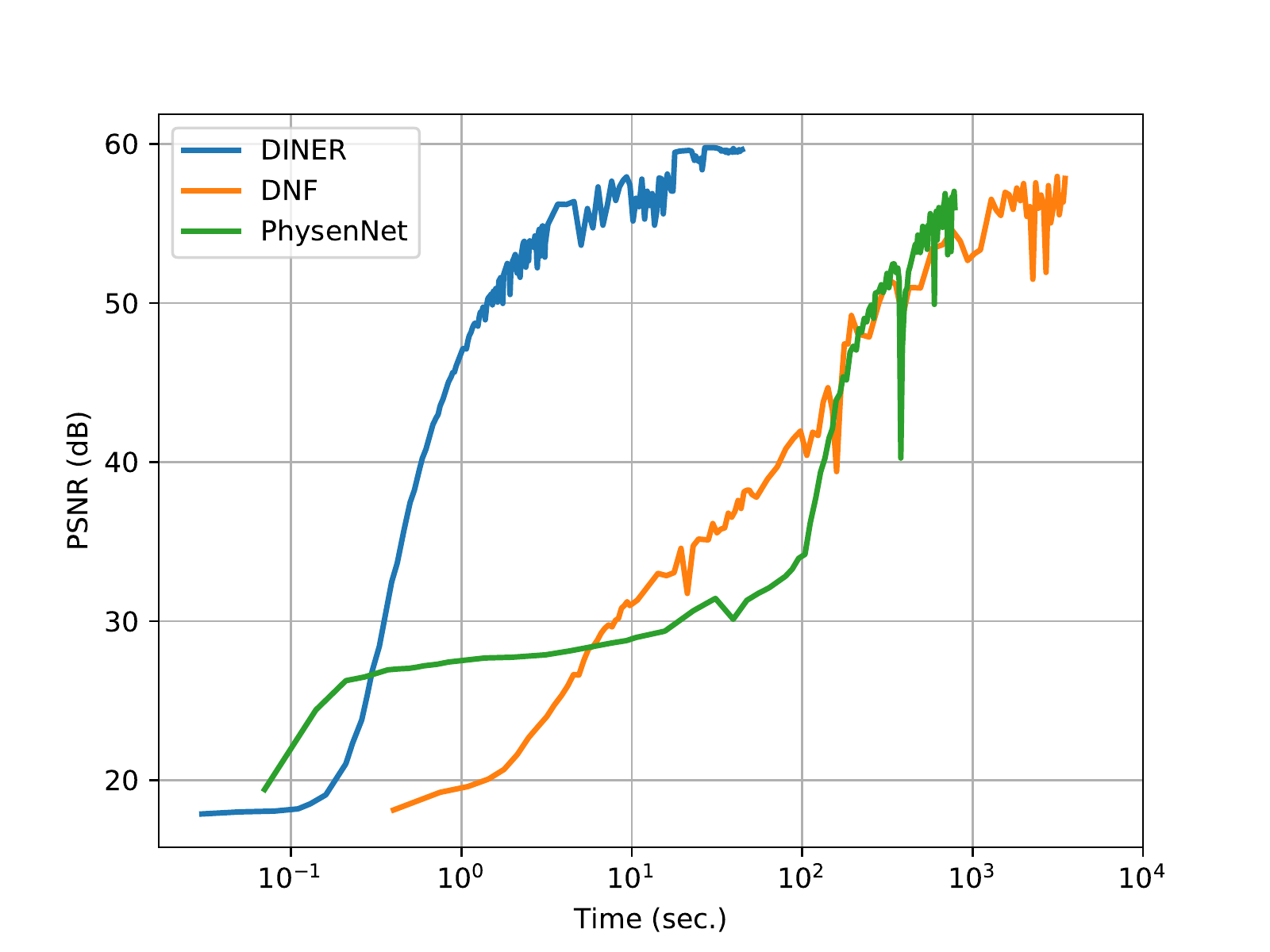}
  \caption{PSNR of reconstructed measurements over training time on lensless imaging.}
  \label{fig:lensless_PSNR_over_epochs}
\end{figure}

\begin{figure}[t]
  \centering
  \includegraphics[width=\linewidth]{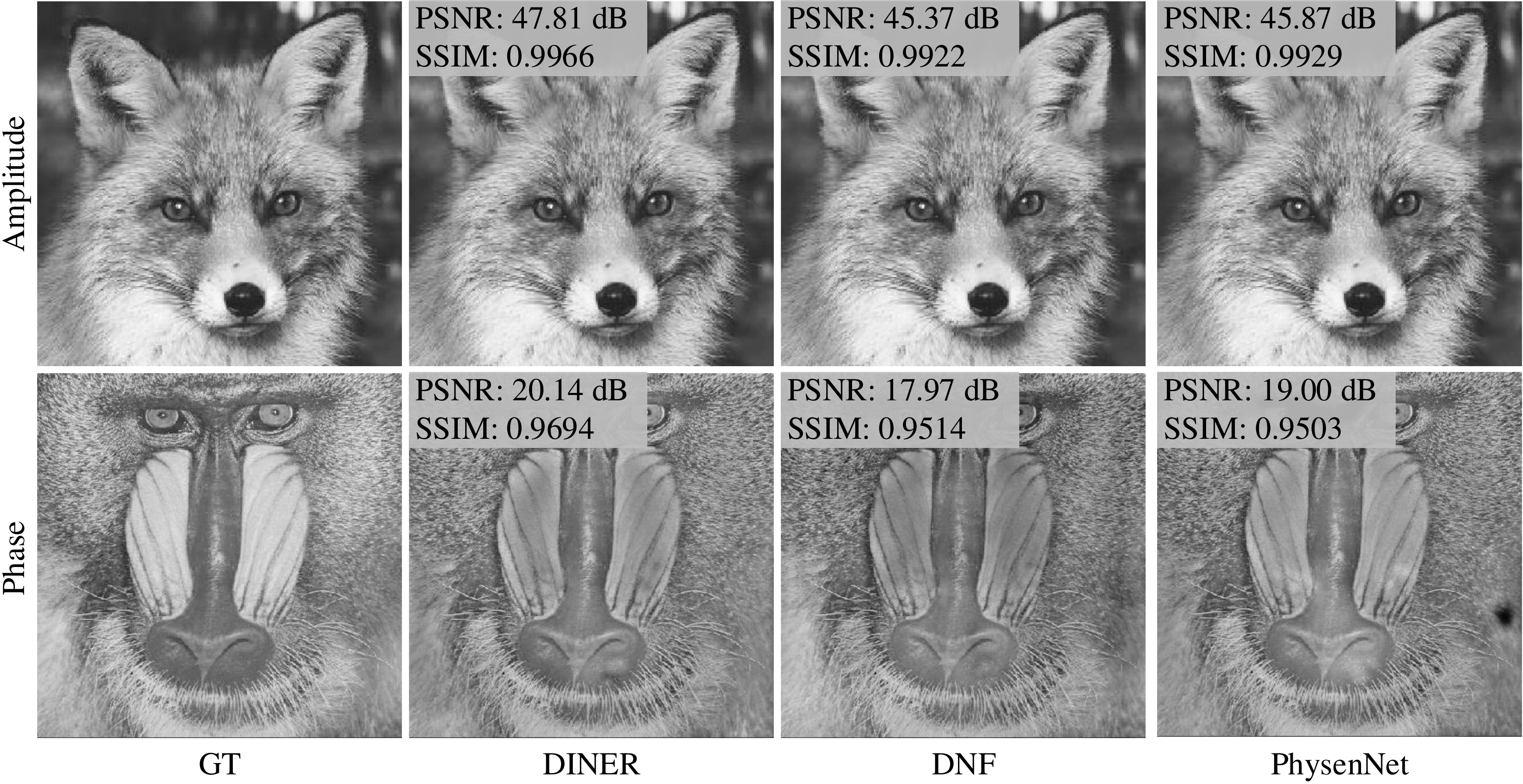}
  \caption{Comparisons on lensless imaging.}
  \label{fig:cmp_lensless}
\end{figure}

\subsection{Phase Recovery in Lensless Imaging}
Lensless imaging~\cite{ozcan2016lensless} observes specimen in a very close distance without any optical lens. By directly recording the diffractive measurements, it provides the advantage of wide field of view observation and has become an attractive microscopic technique~\cite{zhou2020wirtinger} for analyzing the properties of the specimen. We take the classic multi-height lensless imaging as an example, where $N$ measurements $\{I_z\}_{z=z_1}^{z_N}$ are captured under different specimen-to-sensor distances $z$ for the specimen's amplitude and phase imaging recovery. In multi-height lensless imaging, $I_{z}$ could be modeled as applying Fresnel propagation to the complex field $O(x,y)$ of the specimen, \textit{i.e.},
\begin{equation}\footnotesize
\label{eqn:inverse_lensless}
    I_z = |PSF_{z}*[P(x,y)\cdot O(x,y)]|^2,
\end{equation}
where $P(x,y)$ is the illumination pattern, $PSF_{z}$ is the point spread function of Fresnel propagation over distance $z$ between the specimen and the sensor.

We model $O(x,y)$ using the proposed method with the SIREN backbone and the network size is $2\times 64$. The loss function is built by comparing the measurements with the results from applying the Eqn.~\ref{eqn:inverse_lensless} to the network output. We compare our method with the current SOTAs, \textit{i.e.}, the diffractive neural field (DNF)~\cite{zhu2022dnf} and the PhysenNet~\cite{wang2020phase}. Due to the lack of the ground truth in real images, we only provide comparisons on the open-source synthetic data~\cite{zhu2022dnf} here, please refer the supplementary material for more qualitative comparisons on real data.

Fig.~\ref{fig:lensless_PSNR_over_epochs} shows the PSNR of  reconstructed measurements over training time. The proposed method has $18\times$ and $80\times$ advantages on the convergence speed over the PhysenNet and DNF, respectively. Fig.~\ref{fig:cmp_lensless} provides qualitative comparisons. Although only $1.5\%$ network parameters ($2\times 64$ \textit{vs.} $8\times 256$) are used, the proposed DINER could provide better results than the DNF thanks to the hash-table. 

\subsection{3D Refractive Index Recovery in Intensity Diffraction Tomography}
The 3D refractive index characterizes the interaction between light and matter within a specimen. It is an endogenous source of optical contrast for imaging specimen without staining or labelling, and plays an important role in many areas, \textit{e.g.} the morphogenesis, cellular pathophysiology, biochemistry~\cite{park2018quantitative}. Intensity diffraction tomography measures the squared amplitude of the light scatted from the specimen at different angles multiple times and has become a popular technique for recovering the 3D refractive index. 

Given the 3D refractive index $\mathbf{n} = (\mathbf{n}_{re} + j\mathbf{n}_{im})$ of a specimen, where $\mathbf{n}_{re}$ and $\mathbf{n}_{im}$ are the real and imaginary parts of the specimen's refractive index, respectively. The forward imaging process of sensor placed at location $\rho$ could be modeled as 
\begin{equation}\footnotesize
\label{eqn:IDT_forward}
I_{\rho} = \mathbf{A}_{\rho}\Delta \epsilon,
\end{equation} 
where $\mathbf{A}_{\rho}$ records the sample-intensity mapping with the illuminations. $\Delta \epsilon=\Delta \epsilon_{re} + j\Delta \epsilon_{im}$ is the complex-valued permittivity contrast and could be obtained by solving
\begin{equation}\footnotesize
\begin{aligned}
&\mathbf{n}_{\mathrm{re}}=\sqrt{\frac{1}{2}\left(\left(\mathbf{n}_0^2+\Delta \epsilon_{\mathrm{re}}\right)+\sqrt{\left(\mathbf{n}_0^2+\Delta \epsilon_{\mathrm{re}}\right)^2+\Delta \epsilon_{\mathrm{im}}^2}\right)} \\
&\mathbf{n}_{\mathrm{im}}=\frac{\Delta \epsilon_{\mathrm{im}}}{2 \cdot \mathbf{n}_{\mathrm{re}}}
\end{aligned},
\end{equation}
where $\mathbf{n}_0$ is the refractive index of the background medium.

We model the $(\Delta \epsilon_{re}, \Delta \epsilon_{im})$ using the DINER with network size $2\times 64$. We compare our method with the SOTA, \textit{i.e.}, DeCAF~\cite{liu2022recovery} which uses a combination of the standard MLP structure with network size $10\times 208$ and positional+radial encodings. 

\begin{figure}[t]
  \centering
  \includegraphics[width=\linewidth]{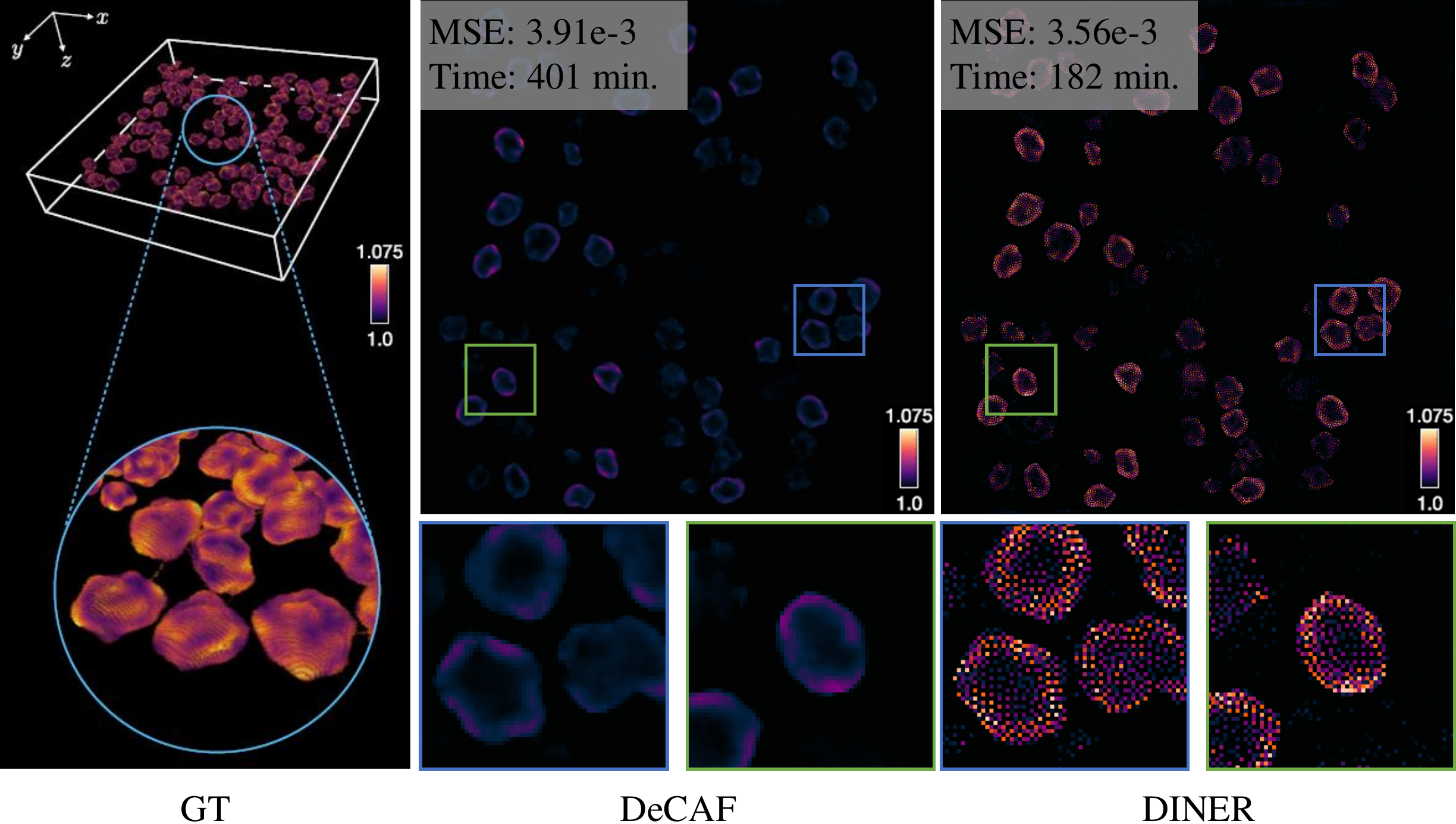}
  \caption{Comparisons on 3D refractive index recovery. DINER takes less training time and could reconstruct more surface details of the Granulocyte.}
  \label{fig:cmp_RI}
\end{figure}

Fig.~\ref{fig:cmp_RI} compares our method with the DeCAF on the 3D Granulocyte Phantom data using the Fiji software~\cite{schindelin2012fiji}. Because the ground truth is not open-source until the submission, we take a screenshot of the ground truth in their paper as a reference. The MSE values labelled in Fig.~\ref{fig:cmp_RI} are computed by comparing the reconstructed measurements using the network output and the real measurements. Since the hash-table could map a high-frequency signal in a low-frequency way, our results provide more details on the surface of the Granulocyte. While the surface of the Granulocyte is over-smoothed in the results of the DeCAF since the PE+MLP could not accurately model the high-frequency components (Please refer to the supplemented video on different heights for better comparison).


\begin{figure}[t]
  \centering
  \includegraphics[width=\linewidth]{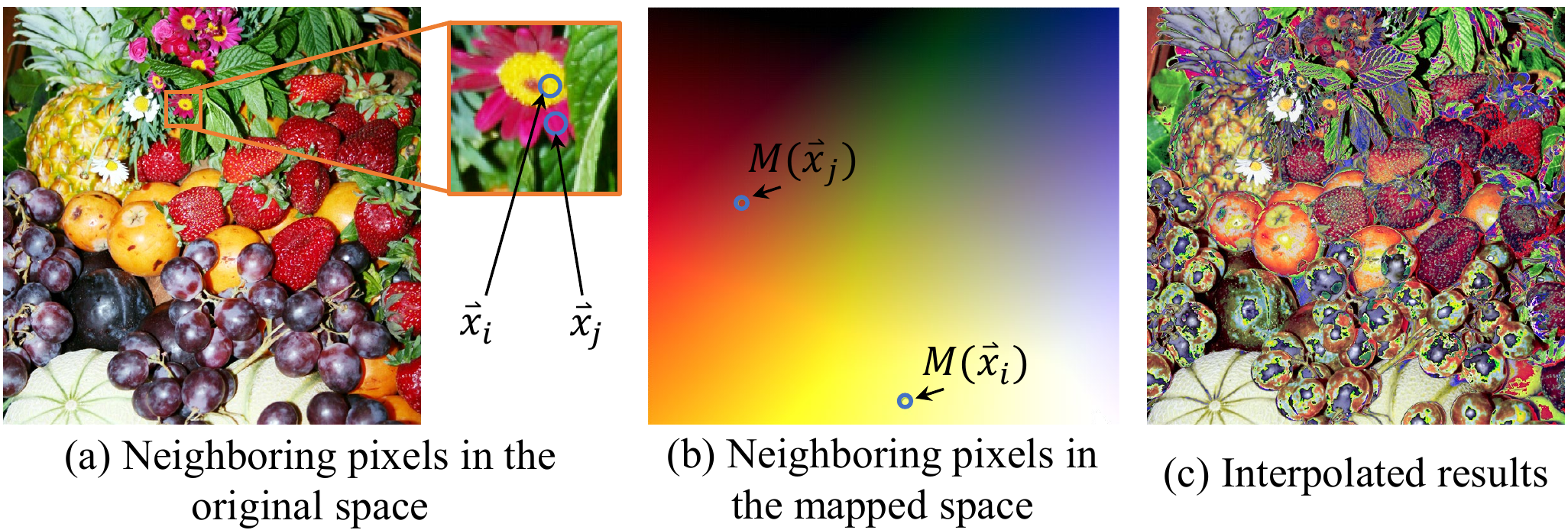}
  \caption{Analysis of feeding interpolated hash-key to the DINER. (a) Two neighboring coordinates $\vec{x}_i$ and $\vec{x}_j$ are labelled in the original image. (b) The distance between the mapped coordinates $\mathcal{HM}(\vec{x}_i)$ and $\mathcal{HM}(\vec{x}_j)$ is larger than the one in the original space. (c) Results by feeding the interpolated mapped coordinates to the trained MLP.}
  \label{fig:limitation}
\end{figure}

\subsection{Discussion}
The aforementioned experiments are all focused on discrete signals. To query an unseen coordinate in a continuous signal, 
it is suggested to apply a post-interpolation operation to the network output instead of feeding interpolated hash-key to the network (see Fig. \ref{fig:limitation}), such as the exploration of Plenoxels which interpolates the grid of density and harmonic coefficients~\cite{fridovich2022plenoxels} instead of feeding unseen position and direction coordinates to the network directly~\cite{mildenhall2020nerf}.


\section{Conclusions}
In this work, we have proposed the DINER which could greatly improve the accuracy of current INR backbones by introducing an additional hash-table. We have pointed out that the performance of INR for representing a signal is determined by the arrangement order of elements in it. The proposed DINER could map the input discrete signal into a low-frequency one, which is invariant if only the arrangement order changes while the histogram of attributes is not changed. For this reason, the accuracy of different INR backbones could be greatly improved. Extensive experiments have verified the high accuracy and efficiency of the proposed DINER for tasks of signal fitting and inverse problem optimization.

However, the current DINER could only process discrete signals. In the future, we will focus on continuous mapping methods instead of discrete hash-table-based mapping to extend the advantages for continuous signals such as the signed distance function~\cite{park2019deepsdf}.




{\small
\bibliographystyle{ieee_fullname}
\bibliography{main}
}

\end{document}